\documentstyle [12pt,epsf] {article}

\catcode`@=12
\newcommand{\beq}{\begin{equation}}
\newcommand{\eeq}{\end{equation}} 
\newcommand{\bea}{\begin{eqnarray}}
\newcommand{\eea}{\end{eqnarray}} 
\newcommand{\bers}{\begin{eqnarray*}}
\newcommand{\eers}{\end{eqnarray*}}
\newcommand{\bt}{\begin{itemize}} 
\newcommand{\et}{\end{itemize}} 

\def\sss{\scriptscriptstyle}
\begin{document} 
\vspace{0.5in} 
\oddsidemargin -.375in 
\newcount\sectionnumber 
\sectionnumber=0 
\def\bra#1{\left\langle #1\right|} 
\def\ket#1{\left| #1\right\rangle} 
\def\be{\begin{equation}} 
\def\ee{\end{equation}} 
\thispagestyle{empty} 
\begin{flushright}  
UdeM-GPP-TH-02-103\\
\end{flushright}
\vskip0.5truecm

\begin{center} 

{\large \bf
\centerline{R-parity-violating SUSY and CP violation in
$B \rightarrow \phi K_S$}}
\vspace*{1.0cm}
{\large 
  Alakabha Datta\footnote{email: datta@lps.umontreal.ca} } \vskip0.3cm
{\it  Laboratoire Ren\'e J.-A. L\'evesque, Universit\'e de
  Montr\'eal,} \\
{\it C.P. 6128, succ.\ centre-ville, Montr\'eal, QC, Canada H3C 3J7} \\
\vskip0.5cm
\bigskip
(\today)
\vskip0.5cm
{\Large Abstract\\}
\vskip3truemm
\parbox[t]{\textwidth} {
Recent measurements of CP asymmetry in $B \rightarrow \phi K_S$
appear to be inconsistent with Standard Model expectations.
We explore the effect of
R-parity-violating SUSY to understand the data.}
\end{center}
\thispagestyle{empty}
\newpage
\setcounter{page}{1}
\baselineskip=14pt

\section{Introduction}

In the standard model (SM), CP violation is
due to the presence of phases in the Cabibbo-Kobayashi-Maskawa (CKM)
quark mixing matrix. 
The SM predicts
large CP-violation in $B$ decays \cite{CPreview}
and the $B$-factories BaBar and Belle will test the SM
explanation of CP violation. 
One of the CP phases of the CKM unitarity triangle has been
already
measured: $\sin 2\beta = 0.78 \pm 0.08$ \cite{betameas}, which is
consistent with the SM.

The goal of the $B$- factories is not only to to test the Standard Model (SM)
picture of CP violation but  also to discover evidence of new physics.
Decays that get significant contributions from penguins are most likely
to be affected by new physics \cite{Gross}. In particular the decay
$B \rightarrow \phi K_S$ is very interesting because it is pure penguin
and is dominated by a single amplitude in the SM. Hence this decay can be
used to measure $\sin(2 \beta)$ and if this measurement is found to disagree
with $\sin(2 \beta)$ from other measurements, like $B \rightarrow J/\psi K_S$,
then it will be a clear sign of new physics \cite{London}.

There have been recent
reported measurements of CP asymmetries in $B \to \Phi K_S$ decays
by BaBar \cite{babarphi} 
\begin{eqnarray}
\sin (2 \beta (\Phi K_S))_{BaBar}=-0.19 ^{+0.52}_{-0.50} \pm 0.09
\label{babar} 
\end{eqnarray}
and Belle \cite{bellephi}
\begin{eqnarray}
\sin (2 \beta (\Phi K_S))_{Belle}=-0.73 \pm 0.64 \pm 0.18
\label{belle}
\end{eqnarray}
Combining the two measurements and adding the
errors  in quadrature one obtains
\begin{equation}
\label{eq:ave}
\sin (2 \beta (\Phi K_S))_{ave}=-0.39 \pm 0.41 \; .
\end{equation} 
This result appears to be inconsistent with SM prediction as
$\sin(2 \beta)$  from $B \rightarrow J/\psi K_s$ should agree
with $\sin(2 \beta)$  from $B \rightarrow \phi K_s$  up to
0($\lambda^2$) with $\lambda \sim 0.2$. However, the measurements
presented above seem to indicate a 2.8 $\sigma$ deviation
 from SM expectation and have led to speculations about
evidence of new physics \cite{Hiller}.
In this paper we study the effect of R-parity violating SUSY
for the decay $B \rightarrow \phi K_S$ and show that
the present measurements of $\sin(2 \beta)$ can be easily accommodated
in the presence of R-parity violating SUSY.
\section{R-parity breaking SUSY and $B \rightarrow \phi K_S$}
In supersymmetric models, $R$-parity invariance is often imposed on
the Lagrangian in order to maintain the separate conservation of
baryon number and lepton number. 
The $R$-parity of a field with spin $S$,
baryon number $B$ and lepton number $L$ is defined to be
\beq
R=(-1)^{2S+3B+L} ~.
\eeq
$R$ is $+1$ for all the SM particles and $-1$ for all the
supersymmetric particles. 

The presence of $R$-parity conservation
implies that super particles must be produced in
pairs in collider experiments and the lightest super particle (LSP)
must be absolutely stable. The LSP therefore provides a good candidate
for cold dark matter.
There is, however, no compelling
theoretical motivation, such as gauge invariance, to impose 
R-parity conservation. 

The most general superpotential of the
MSSM, consistent with $SU(3)\times SU(2)\times U(1)$ gauge symmetry
and supersymmetry, can be written as
\beq
{\cal W}={\cal W}_R+{\cal W}_{\not \! R}~,
\eeq
where ${\cal W}_R$ is the $R$-parity conserving piece, and ${\cal
W}_{\not \! R}$ breaks $R$-parity. They are given by
\bea
{\cal W}_R&=&h_{ij}L_iH_2E_j^c+h_{ij}^{\prime}Q_iH_2D_j^c
             +h_{ij}^{\prime\prime}Q_iH_1U_j^c ~,\\ \label{RV}
{\cal W}_{\not \! R}&=&\frac{1}{2}\lambda_{[ij]k}L_iL_jE_k^c
+\lambda_{ijk}^{\prime}L_iQ_jD_k^c
             +\frac{1}{2}\lambda_{i[jk]}^{\prime\prime}
U_i^cD_j^cD_k^c+\mu_iL_iH_2 ~.
\label{lag}
\eea
Here $L_i(Q_i)$ and $E_i(U_i,D_i)$ are the left-handed lepton (quark)
doublet and lepton (quark) singlet chiral superfields, where $i,j,k$
are generation indices and $c$ denotes a charge conjugate field.
$H_{1,2}$ are the chiral superfields representing the two Higgs
doublets.

The
$\lambda$ and $\lambda^{\prime}$ couplings in
[Eq.~(\ref{RV})],
violate lepton number
conservation, while the $\lambda^{\prime\prime}$ couplings violate
baryon number conservation.  
There are  27
$\lambda^{\prime}$-type couplings and 9 each of the $\lambda$ and
$\lambda^{\prime \prime}$ couplings as
$\lambda_{[ij]k}$ is antisymmetric in the
first two indices and $\lambda^{\prime\prime}_{i[jk]}$ is
antisymmetric in the last two indices. 
The non-observation of proton decay imposes very
stringent conditions on the simultaneous presence
of both the baryon-number and lepton-number violating terms 
in the Lagrangian \cite{Proton}.
It is therefore customary to assume 
the existence of either $L$-violating couplings
or $B$-violating couplings, but not both. The terms proportional to
$\lambda$ are not relevant to our present discussion and will not be
considered further.

The antisymmetry of the $B$-violating couplings,
$\lambda^{\prime\prime}_{i[jk]}$ in the last two indices implies that
there are no operators that can generate the $b \to s \bar{s} s$
transition, and hence cannot contribute to $B \rightarrow \phi K_S$
at least at the tree level.

We now turn to the $L$-violating couplings. In terms of four-component 
Dirac spinors, these are given by \cite{DatXin}
\bea
{\cal L}_{\lambda^{\prime}}&=&-\lambda^{\prime}_{ijk}
\left [\tilde \nu^i_L\bar d^k_R d^j_L+\tilde d^j_L\bar d^k_R\nu^i_L
       +(\tilde d^k_R)^*(\bar \nu^i_L)^c d^j_L\right.\nonumber\\
& &\hspace{1.5cm} \left. -\tilde e^i_L\bar d^k_R u^j_L
       -\tilde u^j_L\bar d^k_R e^i_L
       -(\tilde d^k_R)^*(\bar e^i_L)^c u^j_L\right ]+h.c.\
\label{Lviolating}
\eea
{}For  the $b \to s {\bar{s}}s $ transition, the relevant
Lagrangian is
\beq
L_{eff} = \frac{\lambda^{\prime}_{i32} \lambda^{\prime*}_{i22}} { m_{
\widetilde{\nu}_i}^2} \bar s \gamma_R s \, \bar {s}\gamma_L b+
\frac{\lambda^{\prime}_{i22} \lambda^{\prime*}_{i23}} { m_{
\widetilde{\nu}_i}^2} \bar s \gamma_L s \, \bar {s}\gamma_R b ~,
\eeq
where $\gamma_{R L}= {(1 \pm \gamma_5) \over 2}$.
There are a variety of sources which bound the above couplings
\cite{Rpreview} but the present bounds are fairly weak and the
contribution from the L-violating couplings can significantly affect
$B \rightarrow \phi K_S$ measurements.

In the SM, the amplitude for $B \to \phi K_S$, can be written within factorization as{\footnote{ This decay has been recently studied in QCD factorization
in Ref\cite{Cheng}.}
\bea
A_{\sss \phi K_S}^{\sss SM, } & = 
& -{G_{\sss F} \over \sqrt{2}}
 V_{tb}V_{ts}^*\left[ a_3^t+ a_4^t + a_5^t -\frac{1}{2}a_7^t
 -\frac{1}{2}a_9^t -\frac{1}{2}a_{10}^t \right. \nonumber\\ 
& & \hskip20truemm \left.  -a_3^c- a_4^c - a_5^c +\frac{1}{2}a_7^c
+\frac{1}{2}a_9^c +\frac{1}{2}a_{10}^c \right]Z, \nonumber\\
Z &= & 2f_{\phi}m_{\phi}F_{BK}(m_\phi^2) \varepsilon^*\cdot p_B,
\label{SM}
\eea
where $f_{\phi}$ 
is the $\phi$ decay constant, $F_{BK}$ 
is the $B \rightarrow K$
semileptonic form factor. The $a_i^{t,c}$ are the 
usual combination of Wilson's coefficient
in the effective Hamiltonian. For $a_i$ as well as the quark masses 
we use the values used  
in Ref\cite{KX}. 
The R-parity contribution can be written as,
\bea
A_{\sss \phi K_S}^{\not \! R}  
& =  & (X_1+X_2) Z  \nonumber\\
X_1 &=&-\frac{\lambda^{\prime}_{i32} \lambda^{\prime*}_{i22}} {24 m_{
\widetilde{\nu}_i}^2}\nonumber\\
X_2 & = & -\frac{\lambda^{\prime}_{i22} \lambda^{\prime*}_{i23}} {24 m_{
\widetilde{\nu}_i}^2}\
\label{R}
\eea
\bigskip
\noindent
We  write
\beq
X_1+X_2=\frac{X}{12M^2}e^{i\phi}
\eeq
 where $\phi$ is the weak phase
in the R-parity violating couplings and M is some mass scale
with $M \sim m_{\widetilde{\nu}_i}$.
We require  $|X|$ to be such that 
$|A_{\sss \phi K_S}^{\sss SM, }|$  and
$|A_{\sss \phi K_S}^{\not \! R}|$ 
are of the same size, which then fixes
$|X| \sim 1.5 \times 10^{-3}$ for $M=100$ GeV
 which is  smaller than the existing 
constraints on $|X|$ from Ref\cite{Rpreview}.

We can now calculate $\sin(2 \beta)_{eff}$ 
from
\bea
\sin(2 \beta)_{eff} & = &-{2Im[\lambda_f] \over (1+|\lambda_f|^2)}\nonumber\\
\lambda_f & = &  e^{-2i\beta}{\bar{A} \over A}\
\eea
where $A=A_{\sss \phi K_S}^{\sss SM, }+
A_{\sss \phi K_S}^{\not \! R}$ and $\bar{A}$ is the amplitude for the 
CP-conjugate process. 
Note that from Eq. \ref{SM} and Eq. \ref{R}
$\sin(2 \beta)_{eff}$  is independent of $Z$ and hence 
free from uncertainties in the form factor and decay constants.
It is also possible that non factorizable effects may be less important in
$\sin(2 \beta)_{eff}$ as we are taking ratios of amplitudes.
In Fig. 1 we plot $\sin(2 \beta)_{eff}$  versus the phase $\phi$ and it is 
clear from the figure that the present measurements in Eq. \ref{babar} and
Eq. \ref{belle} can be easily explained.
\begin{figure}[htb] 
   \centerline{\epsfysize 4.2 truein \epsfbox{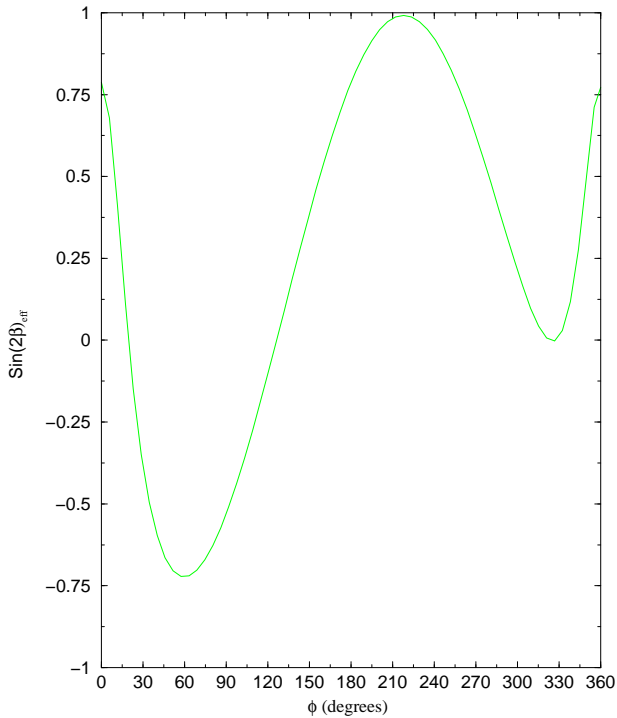}} 
   \caption{$Sin(2 \beta)_{eff}$ versus $\phi$}
\label{phiK.ps} 
\end{figure} 

We now turn to the calculation of branching ratios and the direct CP asymmetry.
The measured branching ratio for  
$(B \to \Phi K^0)$ is $(8.1 ^{+3.1}_{-2.5} \pm 0.8) \times 
10^{-6}$ \cite{PDG2002} while
Belle  measures  a value for the direct CP asymmetry, i.e.~the cosine term
$C=-0.56 \pm 0.41 \pm 0.12$ \cite{bellephi} which is consistent with zero.
The calculation of the branching ratio as well as the direct CP asymmetry is
difficult and suffers from hadronic uncertainties even within factorization.
The branching ratio, within naive factorization, depends on the
 form factors and the $\phi$ decay constants. The uncertainties in these 
quantities can easily change the predicted branching ratio by a factor of 2 
or so. The direct CP asymmetry could potentially be large as there are
two interfering amplitude of the same size. However the direct CP asymmetry 
crucially depends on the strong phase which can be 
perturbatively generated through tree level rescattering in factorization.
The size of this strong phase, in this case, 
depends on the charm quark mass as well and the gluon momentum in 
the penguin diagram.
Using the values of the formfactor $F_{BK}=$ 0.38 
and the $\phi$ decay constant $f_{\phi}=$ 0.237 \cite{Cheng} 
and taking a typical value for the phase $\phi=1.5$ radians(86 degrees) 
we obtain the branching ratio for
$(B \to \Phi K^0)=9.5 \times 10^{-6}$, $\sin{2 \beta}=-0.57$ 
and
the direct CP asymmetry $ \sim 35$ \%. We have used $m_c=1.4$ GeV, 
$m_b=5$ GeV and the gluon momentum $q^2=m_b^2/2$ to obtain these numbers. 
We would like to stress here that 
even in the absence of the strong phase one can still easily accommodate
the data for $\sin{2 \beta}$ in $ B \rightarrow \phi K_S$. 
If in fact the strong phases are small 
one could look for T- violating correlation in
$B \rightarrow \phi K^*$
or in the corresponding  $\Lambda_b$ decays\cite{BDL}.

We also point out that the R-parity violating operator for 
$ b \rightarrow s {\bar{s}}s$ 
is not the related to the $ b \rightarrow s {\bar{u}}(\bar{d})u(d)$, unlike
some models of new physics. Hence R-parity violating effects in
$B \rightarrow \phi K_S$ can be very different 
than in $B \to K \pi$ for example which is a 
$ b \rightarrow s {\bar{u}}u$ transition.
However the new R-parity violating SUSY generated
 $ b \rightarrow s {\bar{s}}s$ operators will affect decays like
$B \rightarrow K(K^*) \eta(\eta^{\prime})$, 
$\Lambda_b \rightarrow \Lambda \eta(\eta^{\prime})$,   
$\Lambda_b \rightarrow \Lambda \phi$ etc \cite{BDLN}.

In conclusion, recent
 measurements of CP asymmetry in $B \rightarrow \phi K_S$
appear to be inconsistent with Standard Model expectations.
We show that the effect of
R-parity-violating SUSY can easily accommodate the data.
   
{\bf Acknowledgements}:
This work was financially supported by NSERC of Canada.


\end{document}